\documentclass[runningheads]{llncs}
\usepackage[T1]{fontenc}
\usepackage{graphicx}
\usepackage{xcolor} 
\usepackage{cite}         
\usepackage{amsmath,amssymb,amsfonts}  
\usepackage{algorithmic}  
\usepackage{hyperref}     
\usepackage{xurl} 
\usepackage{array}        
\usepackage{booktabs}     
\usepackage{tabularx}     
\usepackage{orcidlink}
\usepackage{longtable}
\usepackage{float}
\begin{document}
\title{Impact and Implications of Generative AI for Enterprise Architects in Agile Environments: A Systematic Literature Review}
\titlerunning{Impact and Implications of GenAI on Enterprise Architects}
%
\author{Stefan Julian Kooy\,\orcidlink{0009-0008-7094-1486}\and
Jean Paul Sebastian Piest\,\orcidlink{0000-0002-0995-6813} \and
Rob Henk Bemthuis\,\orcidlink{0000-0003-2791-6070}}
\authorrunning{S.J. Kooy et al.}
%
\institute{University of Twente, Drienerlolaan 5, 7522 NB, The Netherlands\\
\email{s.j.kooy@student.utwente.nl}, 
\email{\{j.p.s.piest,r.h.bemthuis\}@utwente.nl}}
\maketitle              
\begin{abstract} 
Generative AI (GenAI) is reshaping enterprise architecture work in agile software organizations, yet evidence on its effects remains scattered. We report a systematic literature review (SLR), following established SLR protocols of Kitchenham and PRISMA, of 1,697 records, yielding 33 studies across enterprise, solution, domain, business, and IT architect roles. GenAI most consistently supports (i) design ideation and trade-off exploration; (ii) rapid creation and refinement of artifacts (e.g., code, models, documentation); and (iii) architectural decision support and knowledge retrieval. Reported risks include opacity and bias, contextually incorrect outputs leading to rework, privacy and compliance concerns, and social loafing. We also identify emerging skills and competencies, including prompt engineering, model evaluation, and professional oversight, and organizational enablers around readiness and adaptive governance. The review contributes with (1) a mapping of GenAI use cases and risks in agile architecting, (2) implications for capability building and governance, and (3) an initial research agenda on human-AI collaboration in architecture. Overall, the findings inform responsible adoption of GenAI that accelerates digital transformation while safeguarding architectural integrity. 

\end{abstract}

\section{Introduction}
Organizations face persistent challenges in developing and delivering digital products and services. Agile methods such as Scrum and frameworks like the Scaled Agile Framework (SAFe) have reshaped how teams conceive, build, and deploy solutions, emphasizing flexibility and rapid feedback to meet changing market needs~\cite{kalenda2018scaling}. In parallel, the adoption of generative artificial intelligence (GenAI) is accelerating across manufacturing, healthcare, finance, telecommunications, and the public sector~\cite{McKinsey2023AI,rashid2024ai,Nah2023, BrightEtAl2024_GenerativeAI_PublicSector}. GenAI refers to the class of Artificial Intelligence (AI) models that emulate the structure and characteristics of input data to generate synthetic content, such as text, images, audio, or video, rather than merely classifying or predicting existing data~\cite{NIST2024}. GenAI is increasingly used to streamline operations, support customer interaction, and enable new forms of innovation, making it a practical enabler for modernizing industrial practice. 

Enterprise Architecture (EA) offers guidance for aligning digital solutions, from cloud-native platforms to data, Internet of Things (IoT), and Artificial Intelligence (AI), with strategic objectives. In this paper, “architects” includes enterprise, solution, domain, business, and IT architects. Practitioners frequently rely on frameworks such as TOGAF or Zachman to support business-IT alignment and strategic decision making. However, traditional approaches can fall short in agile settings due to “big design upfront” practices and slower feedback cycles~\cite{waterman2015agile}. Prior work indicates that AI can improve decision making within established architecture frameworks, a foundation for organizational agility~\cite{nafei2016organizational}, yet integrating AI into these frameworks remains challenging~\cite{Bakar2024}. 

This paper examines how GenAI affects the role, skills, and responsibilities of architects in large-scale agile software development. We conduct a systematic literature review (SLR) to assess how GenAI may address current challenges and to identify the capabilities relevant to architecture work (e.g., generating design alternatives, accelerating documentation, supporting architectural decisions). We also consider potential risks and trade-offs. Our goal is to clarify whether GenAI can contribute to architectural agility and operational efficiency, and under what conditions. 

The remainder of the paper is organized as follows. Section~\ref{section:background} outlines the background on EA, agile methodologies, and GenAI. Section~\ref{section:methodology} introduces the methodology. Section~\ref{section:results} presents the key results and findings of the SLR. Section~\ref{section:discussion} discusses the implications of the results and findings, and Section~\ref{section:conclusion} concludes with a summary, limitations, and directions for future research. 

\section{Background}
\label{section:background}
This section defines key concepts: EA as the structure for aligning business and technology, common EA frameworks, agile methods for software delivery, and GenAI.

\subsection{Enterprise Architecture}
EA provides a structured approach to designing and governing systems, technologies, and processes so they align with business objectives and operate effectively in dynamic environments~\cite{Lankhorst2017,strano2007enterprise}. EA spans multiple architectural viewpoints (enterprise, business, IT, domain, solution) that share key responsibilities: stakeholder communication, strategic alignment, risk management, and governance~\cite{gellweiler2020itarch}. As an organizational blueprint, EA helps ensure that technology investments support business goals, improve agility, and address regulatory and cybersecurity concerns while reducing complexity~\cite{strano2007enterprise,gellweiler2020itarch,mulder2023modern}. Standardized processes and consistent documentation also support digital transformation, including the adoption of AI-enabled capabilities~\cite{bughin2024corporate,georgiev2024knowledge}. 

\subsection{Architecture Frameworks}
The Zachman Framework~\cite{Zachman1987,Zachman2000} and The Open Group Architecture Framework (TOGAF)~\cite{TOGAF9_2025} offer widely used approaches for implementing EA. Zachman organizes architectural artifacts across the “What, How, Where, Who, When, Why” perspectives and role-based rows to support comprehensive documentation and cross-functional collaboration. TOGAF's Architecture Development Method (ADM) defines iterative phases for aligning business and IT, managing change, and governing execution. While effective for long-term planning, these phase-driven approaches can conflict with the fast feedback cycles and flexibility emphasized by agile methods~\cite{Rouhani2015}. 

\subsection{Agile Methodologies}
Agile methodologies, such as Scrum, and frameworks, such as SAFe, promote iterative development, short feedback loops, and cross-functional collaboration~\cite{beck2001agile,Waterfall,sassa2023scrum}. Scrum is common in small teams that work in short sprints with frequent stakeholder input and incremental releases. SAFe scales agile practices across the enterprise via “Agile Release Trains” and formal architecture roles (enterprise, solution, and system) that help maintain strategic alignment and architectural integrity~\cite{putta2018adopting,scaledagile_enterprise_architect,scaledagile_solution_architect,scaledagile_system_architect}. Despite these benefits, agile practices can strain traditional EA processes that rely on upfront design and rigid governance~\cite{Rouhani2015}. 

\subsection{Generative AI}
AI aims to build systems that perform tasks requiring human-level capabilities such as pattern recognition, language understanding, and decision making~\cite{williams1983aiintroduction}. Machine learning and deep learning learn from data rather than explicit rules. GenAI focuses on producing new content. Using models such as GANs, VAEs, Transformers, and latent diffusion models, GenAI can generate text, code, and images from learned distributions~\cite{Banh2023,Chahal2019}. 

GenAI may help bridge the structure of EA with the iterative nature of agile work. It can automate routine documentation and augment design ideation, potentially easing common bottlenecks. This review therefore studies GenAI's impact on architectural roles and responsibilities in agile environments, with attention to implications for practice, decision making, and governance.

\section{Methodology}
\label{section:methodology}
We followed Kitchenham's guidelines~\cite{kitchenham2018guidelines} and PRISMA~\cite{PRISMA2021} to plan, conduct, and report this SLR. The sections below describe how we defined the research questions (RQs), designed the search strategy, applied inclusion and exclusion criteria, extracted and synthesized data, and assured quality. The full protocol is available at~\cite{OSF}.

\subsection{Research Questions}
The study examines the impact of GenAI on EA in agile development contexts. Our main research question is: \textit{How does GenAI influence the evolving role, skills, and responsibilities of architects in large-scale agile software development environments?}

To operationalize the main RQ, we formulated five sub-questions: 

\begin{enumerate}
    \item Which technical characteristics of GenAI create opportunities and challenges for architects in agile environments? 
    \item Which current GenAI use cases influence architectural practices and decision making in agile organizations? 
    \item How does GenAI affect traditional roles, tasks, and required skills of architects in agile digital transformations? 
    \item What organizational and technological factors influence the adoption trajectory of GenAI in architecture practices? 
    \item What governance and capability adaptations are needed to integrate GenAI into architecture practices? 
\end{enumerate}

\subsection{Search Strategy}
The search was performed on 17th of February, 2025, in two electronic databases, IEEE Xplore and Scopus. These databases provide complementary coverage of computing research and practitioner-oriented literature. Guided by the RQs, we designed English-language queries addressing (1) GenAI characteristics, (2) architectural roles and skills, (3) agile architecture, and (4) AI governance and adoption. 

\begin{figure}[ht!]
    \centering
    \includegraphics[width=1.0\textwidth]{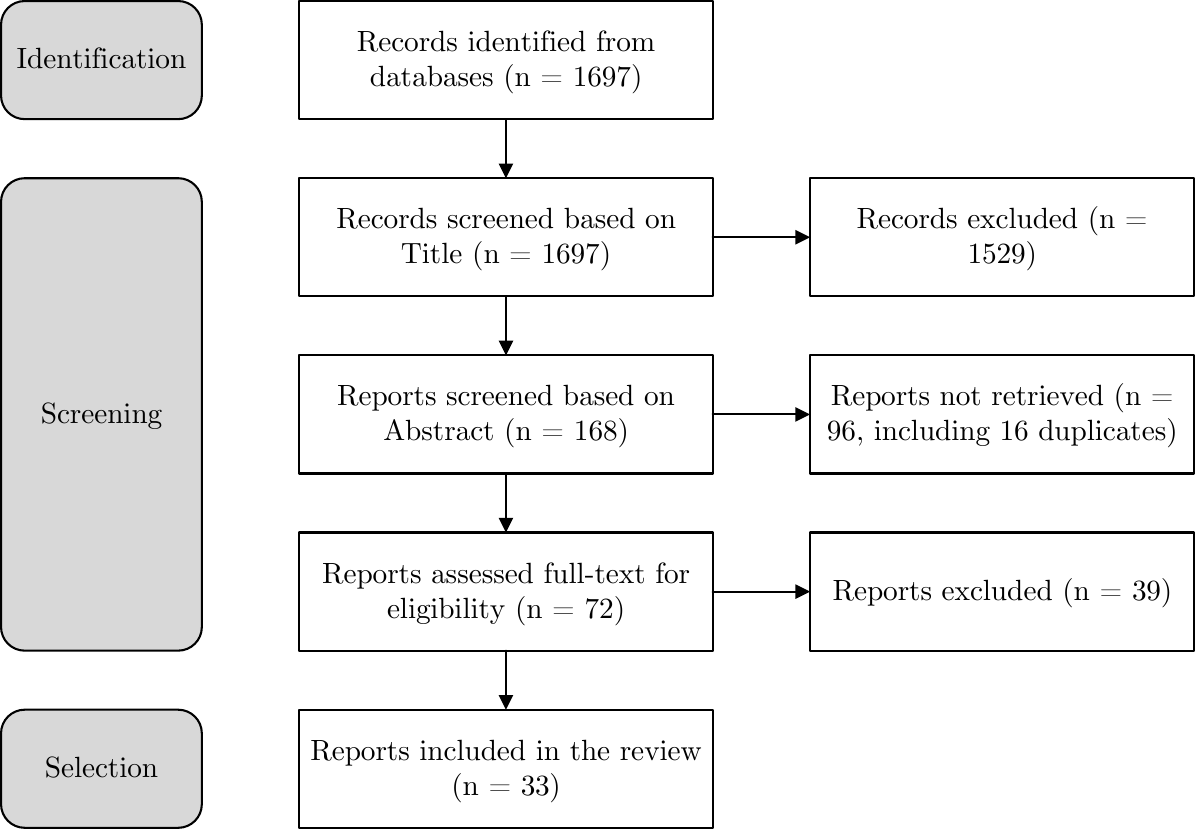}
    \caption{Screening and selection process, as adopted by~\cite{PRISMA2021,kitchenham2018guidelines}.}
    \label{fig:screeningandselectionprocess}
\end{figure}

Four query sets (Table~\ref{tab:searchqueries}) returned 1{,}697 records. Title screening excluded 1{,}529 irrelevant items. Abstract screening removed a further 96 records (including 16 duplicates). We assessed 72 full texts and selected 33 papers for data extraction (Figure~\ref{fig:screeningandselectionprocess}). 

\begin{table}[ht]
\caption{Overview of search queries and results.}
\label{tab:searchqueries}
\centering
\begin{tabular}{|p{1.5cm}|p{7cm}|p{1.5cm}|p{1.5cm}|}
\hline
\textbf{Query} & \textbf{Generic Search String, Specifics In ~\cite{OSF}} & \textbf{Total Found} & \textbf{Selected} \\
\hline
Q1 & \texttt{"Generative AI" AND agile AND (characteristics OR capabilities OR features OR opportunities OR challenges)} & 369 & 11 \\
\hline
Q2 & \texttt{("enterprise architect" OR "domain architect" OR "solution architect" OR "business architect" OR "IT architect") AND (role OR task OR skill)} & 871 & 15 \\
\hline
Q3 & \texttt{("enterprise architect" OR "domain architect" OR "solution architect" OR "business architect" OR "IT architect") AND (agile OR Scrum OR SAFe OR DevOps)} & 390 & 3 \\
\hline
Q4 & \texttt{(("Generative AI" OR "generative artificial intelligence") AND governance) OR ("AI" AND "maturity model")} & 67 & 4 \\
\hline
\end{tabular}
\end{table}

The mapping between RQs and query sets is many-to-many. Each query captures a facet of the topic, with intentional overlap (Table~\ref{tab:dataextractionform}). For example, Q1 targets GenAI characteristics (supporting RQ1); Q2--Q3 emphasize architectural roles, skills, and agile contexts (informing RQ2--RQ3); Q4 focuses on governance and maturity (supporting RQ4--RQ5). We synthesize insights across all queries to answer the five RQs. 

\subsection{Inclusion and Exclusion Criteria}
We defined criteria based on topical relevance, publication type, year, and language (Table~\ref{tab:criteria}). To reflect the recent emergence of GenAI, we applied a stricter recency threshold to GenAI-specific work. 

\begin{table}[ht!]
\caption{Inclusion and exclusion criteria.}\label{tab:criteria}
\centering
\begin{tabular}{|p{6cm}|p{5.5cm}|}
\hline
\textbf{Inclusion Criteria} & \textbf{Exclusion Criteria} \\
\hline
Studies addressing at least one of the RQs 
& Studies focusing on non-GenAI (e.g., rule-based systems, traditional ML, robotics) \\
\hline
Subject areas: computer science, social sciences, engineering, decision science 
& Studies from domains (e.g., medicine, agriculture) lacking relevant GenAI use cases for EA \\
\hline
Publication types: journal articles, SLRs, meta-analyses, conference papers or books 
& Articles published in languages other than English \\
\hline
Published prior to 2015 only if foundational for EA or agile contexts 
& --- \\
\hline
Published in 2015 or later (for studies related to EA, agile methods, or architect roles) 
& --- \\
\hline
Published in 2022 or later (for studies related to GenAI) 
& --- \\
\hline
\end{tabular}
\end{table}

\subsection{Data Extraction and Synthesis}
We extracted data against the RQs (Table~\ref{tab:dataextractionform}). We then performed mapping and thematic analysis to identify cross-study patterns and synthesize findings. 

\begin{table}[ht!] 
\caption{Relevance of studies to research questions.}\label{tab:dataextractionform}
\centering
\begin{tabular}{|c|c|c|c|c|c|c|}
\hline
\textbf{ID} & \textbf{RQ1} & \textbf{RQ2} & \textbf{RQ3} & \textbf{RQ4} & \textbf{RQ5} & \textbf{Ref.} \\
\hline
\ 1 & x & & & & x & \cite{Wang2023} \\
\ 2 & x & x & & & & \cite{Paliwal2024} \\
\ 3 & x & x & & x & & \cite{Hamza2024} \\
\ 4 & x & & & & & \cite{Hagos2024} \\
\ 5 & x & & & & x & \cite{Golda2024} \\
\ 6 & x & x & & x & x & \cite{Alenezi2025} \\
\ 7 & x & & & x & & \cite{DiRocco2025} \\
\ 8 & x & x & & & & \cite{Norheim2024} \\
\ 9 & & & & & x & \cite{mahmoud2025public} \\
\ 10 & x & x & & & & \cite{Handler2024} \\
\ 11 & x & x & & & & \cite{Bahi2024} \\
\ 12 & & x & x & & & \cite{ullrich2021vuca} \\
\ 13 & & & x & & & \cite{kempegowda2020essential} \\
\ 14 & & & x & & & \cite{vandenberg2016roles} \\
\ 15 & & & x & & & \cite{Wagner2016} \\
\ 16 & & x & x & & & \cite{Berg2016decision} \\
\ 17 & x & & x & & & \cite{krishnamurthy2017solution} \\
\ 18 & & & x & & & \cite{Canat2018} \\
\ 19 & & & x & & & \cite{razavian2023odaks} \\
\ 20 & & & & x & & \cite{kaddoumi2022foundational} \\
\ 21 & & & x & & & \cite{kurnia2021enterprise} \\
\ 22 & & & x & & & \cite{gellweiler2019collaboration} \\
\ 23 & & & x & & & \cite{uludag2019expectations} \\
\ 24 & & & x & & & \cite{Breithaupt2021} \\
\ 25 & & & x & & & \cite{gellweiler2020itarch} \\
\ 26 & & & x & & & \cite{gellweiler2022alignment} \\
\ 27 & & & x & & & \cite{uludag2020patterns} \\
\ 28 & & & x & & & \cite{uludag2020supporting} \\
\ 29 & & & x & & & \cite{Guo2021} \\
\ 30 & x & & & x & x & \cite{alkfairy2025strategic} \\
\ 31 & x & & & & x & \cite{amendi2024data} \\
\ 32 & & & & x & x & \cite{nakayama2024balancing} \\
\ 33 & x & & & & & \cite{nidhisree2024generative} \\
\hline
\end{tabular}
\end{table}

\subsection{Quality Assurance}
We assessed methodological rigor, transparency, and potential biases~\cite{OSF}. Overall, the included literature provides a basis for analysis. Many studies use structured methods (e.g., SLRs, case studies, empirical designs) with clear objectives and analytical frameworks. Rigor is strongest in work that follows established guidance (e.g., PRISMA), and data collection and analysis are generally well documented. Some papers exhibit partial outcome reporting or limited methodological detail (e.g.,~\cite{Hamza2024,amendi2024data,Guo2021}). Several conceptual papers provide fewer reflections on assumptions and limitations (e.g.,~\cite{gellweiler2020itarch,Hagos2024}).

\section{Results}
\label{section:results}
We report findings by the five RQs. In total, 33 studies were included. Study methods are summarized in Table~\ref{tab:study_methods}. 

\begin{table}[ht!]
\caption{Study methods of included articles.}\label{tab:study_methods}
\centering
\begin{tabular}{|p{7.2cm}|c|p{3cm}|}
\hline
\textbf{Method} & \textbf{Count (n)} & \textbf{References} \\
\hline
Survey & 3 & \cite{Wang2023, Golda2024, kaddoumi2022foundational} \\
\hline
Literature review & 5 & \cite{Hagos2024, Alenezi2025, Norheim2024, gellweiler2019collaboration, nidhisree2024generative} \\
\hline
Systematic literature / mapping study & 2 & \cite{alkfairy2025strategic, Guo2021} \\
\hline
Case study (single, multiple, or mixed-methods) & 8 & \cite{vandenberg2016roles, Wagner2016, Berg2016decision, Canat2018, Breithaupt2021, uludag2020patterns, uludag2020supporting, uludag2019expectations} \\
\hline
Conceptual / Theoretical / Descriptive analysis & 6 & \cite{Paliwal2024, Handler2024, ullrich2021vuca, kempegowda2020essential, razavian2023odaks, gellweiler2022alignment} \\
\hline
Exploratory / Qualitative / Reflective study & 6 & \cite{Hamza2024, Bahi2024, DiRocco2025, kurnia2021enterprise, krishnamurthy2017solution, nakayama2024balancing} \\
\hline
Quantitative (e.g., PLS-SEM) & 1 & \cite{amendi2024data} \\
\hline
Data / Content analysis & 2 & \cite{mahmoud2025public, gellweiler2020itarch} \\
\hline
\end{tabular}
\end{table}

\subsection{Opportunities and Challenges of GenAI for Architects}
This subsection addresses RQ1 by identifying the technical characteristics of GenAI that present opportunities and challenges for architects operating in agile software development environments. Table~\ref{tab:GenAI_architects} summarizes the main opportunities and challenges. 

\begin{table}[ht]
\caption{Opportunities and challenges of GenAI for architects.}\label{tab:GenAI_architects}
\centering
\begin{tabular}{|p{6cm}|p{5.5cm}|}
\hline
\textbf{Opportunities} & \textbf{Challenges} \\
\hline
Automated generation of code, models, and documentation accelerates delivery~\cite{Hamza2024, Paliwal2024} 
& GenAI systems often function as “black boxes”, limiting transparency and traceability~\cite{Alenezi2025, Hagos2024} \\
\hline
Supports fast design ideation and exploration of alternative solutions~\cite{Paliwal2024} 
& Risk of generating biased, inappropriate, or low-quality outputs~\cite{Norheim2024, Golda2024} \\
\hline
Enhances collaboration (e.g. during agile sprints)~\cite{Hamza2024} 
& Over-reliance on AI may impact architects’ analytical judgment and design skills~\cite{krishnamurthy2017solution} \\
\hline
Improves requirements analysis and specification via NLP capabilities~\cite{Norheim2024} 
& Limited context windows reduce effectiveness in large-scale enterprise modeling~\cite{Hagos2024} \\
\hline
Enables architectural decision support through scenario simulation~\cite{Handler2024, Bahi2024} 
& Raises concerns about security, privacy, and regulatory compliance~\cite{Golda2024} \\
\hline
Assists in retrieving and organizing architectural and technical knowledge~\cite{Hagos2024, DiRocco2025} 
& Lack of standardization in GenAI outputs complicates reuse and interoperability~\cite{DiRocco2025} \\
\hline
\end{tabular}
\end{table}

GenAI introduces capabilities that can reshape architecture work in agile settings. Key opportunities include content generation, design ideation, rapid prototyping, and targeted knowledge retrieval. These align with agile practices such as iterative delivery and frequent feedback. LLMs (e.g., ChatGPT) can accelerate the creation of architectural artifacts (documentation, diagrams, and code) and improve cross-role communication~\cite{Hamza2024,Paliwal2024,DiRocco2025,alkfairy2025strategic}. This, in turn, allows enterprise and solution architects to complete tasks more quickly and align IT capabilities with evolving business strategies, while remaining responsive to stakeholder needs~\cite{amendi2024data, krishnamurthy2017solution}. GenAI also supports architectural decision making by suggesting alternative solutions and enabling comparative evaluations~\cite{Paliwal2024,Handler2024}. 

These gains come with risks. A prominent technical challenge is the opacity of GenAI systems; many function as “black boxes” whose internal reasoning processes are inaccessible~\cite{Wang2023}. This can undermine the transparency, interpretability, and traceability required for sound architectural governance~\cite{Alenezi2025,Hagos2024}. Additionally, concerns exist around biased or low-quality outputs, data privacy risks when using sensitive inputs, and a low output reliability, often stemming from hallucinations, where the model generates plausible yet factually incorrect responses. These challenges raise substantial implications for regulatory compliance and architectural integrity~\cite{DiRocco2025, Norheim2024, nidhisree2024generative, Golda2024}. 

Excessive dependence on GenAI can also erode critical thinking skills among architects, weakening their ability to recognize systemic trade-offs or perform nuanced risk assessments. Furthermore, many LLMs often exhibit limited effectiveness in large-scale enterprise contexts on input length and lack of embedded domain knowledge~\cite{Bahi2024,nidhisree2024generative}. This limitation reduces their utility in end-to-end architectural reasoning~\cite{Hagos2024,Wagner2016, Wang2023}.

\subsection{GenAI Use Cases for Architectural Decision Making}
This section is related to RQ2 and discusses GenAI use cases for decision making.

Our findings suggest that GenAI proves particularly impactful during the early phases of agile development, including ideation, trade-off analysis, and requirements engineering. In these stages, GenAI can facilitate the exploration of alternative design paths, enabling architects to balance competing quality attributes, such as scalability, performance, and maintainability~\cite{Paliwal2024}. This can be valuable in volatile and complex enterprise environments, where decision latency should be minimized~\cite{ullrich2021vuca}.

GenAI also improves team collaboration by simplifying complex technical concepts in accessible terms. By bridging gaps between diverse stakeholders, such as developers, product owners, and business analysts, it strengthens the architect's integrative and coordination roles~\cite{Hamza2024,Berg2016decision}. 

Additional use cases involve classifying, prioritizing, and clustering stakeholder requirements, generating architectural diagrams (e.g., UML), and visualizing dependencies early in the design process~\cite{Alenezi2025,Norheim2024}. These functionalities contribute to faster and more informed architectural decision making. 

Nonetheless, human oversight remains key. Given the probabilistic nature of LLMs, their outputs are neither guaranteed to be optimal nor contextually precise, and studies highlight the importance of validating GenAI-derived suggestions before implementation~\cite{Handler2024,Bahi2024,Hamza2024}. 

\subsection{Evolving Role, Tasks, and Required Skills of the Architect}
Addressing RQ3, this subsection examines how GenAI can alter the roles, tasks, and competencies of enterprise architects in agile environments. 

The capabilities of GenAI influence the professional identity and skill composition of architects in agile enterprises. As discussed in previous sections, GenAI can enable tasks such as natural language-based code generation, real-time feedback, and architectural synthesis. These functionalities alleviate (parts of) the cognitive workload traditionally carried out by architects during design decision making, affecting the architect's role~\cite{Berg2016decision}. Recent research on on-demand architectural knowledge systems supports this view, showing how GenAI can deliver generated suggestions, structured knowledge, and reflective prompts to assist in complex architectural evaluations~\cite{razavian2023odaks}. Consequently, architects increasingly operate as curators and validators of GenAI outputs, rather than primary creators of design artifacts. This shift requires a reconfiguration of core architectural competencies. 

There are many different types of architects. While each type has its own responsibilities, they share overlap in tasks and required competencies~\cite{gellweiler2020itarch}. In addition to foundational skills of architects such as technical expertise, stakeholder communication, strategic thinking, and governance oversight~\cite{ullrich2021vuca, gellweiler2022alignment, kurnia2021enterprise, Wagner2016}, architects increasingly require proficiency in prompt engineering, model governance, professional judgement, and human oversight of AI-generated content. These emerging competencies are aligned with the need for fast validation, continuous iteration, and collaborative judgment in dynamic software environments~\cite{kempegowda2020essential, uludag2020supporting, Guo2021}. 

In large-scale agile programs, architects are expected to engage more closely with development teams and operational workflows~\cite{gellweiler2019collaboration}. Prior research shows that architectural practices are increasingly defined by facilitation, negotiation, and design adjudication across multiple agile teams~\cite{uludag2019expectations, uludag2020patterns}. As GenAI capabilities integrate into these workflows, architects assume multi-faceted roles. Depending on the context, they may act as strategic advisors, technical coaches, reviewers, or even crisis managers~\cite{vandenberg2016roles, uludag2020supporting, Breithaupt2021}. This development expands the architectural role from one of high-level oversight to active participation in lower-level software delivery, which could address persistent communication gaps experienced between higher and lower levels~\cite{Canat2018}. 

While traditionally these roles required extensive experience in abstract modeling, stakeholder communication, and standardization across domains~\cite{krishnamurthy2017solution}, these competencies are now augmented, rather than replaced, by AI-driven assistance. Rather than reducing the architect's contribution, GenAI supports their role in contextualizing AI outputs, aligning them with enterprise objectives, and supporting collaborative decision making in agile environments.

\subsection{Factors Influencing GenAI Adoption Trajectory}
In response to RQ4, this subsection identifies the organizational and technological factors that affect GenAI adoption in architectural practices.

Key enablers include targeted education programs, GenAI literacy, and organizational culture. Absent sufficient investments in upskilling (particularly in areas such as prompt engineering, GenAI evaluation, and responsible AI use), organizations may struggle to realize the promised productivity gains~\cite{Alenezi2025,DiRocco2025}. These training efforts are especially important in agile contexts, where architects must adapt to fast feedback loops and evolving stakeholder needs. Additionally, two behavioral challenges are identified during human-AI collaboration: a steep learning curve for mastering GenAI tools, and social loafing, where individuals may rely too much on AI and reduce their own effort~\cite{Hamza2024}. These findings reinforce the importance of comprehensive education and awareness programs. 

Governance maturity also plays a key role. Organizations with formalized AI policies, ethical oversight, and traceability mechanisms are more likely to integrate GenAI into architecture workflows effectively~\cite{nakayama2024balancing}. A maturity model could range from “Initial” to “Optimizing”, where higher levels are characterized by continuous monitoring, defined responsibilities, and established safeguards~\cite{nakayama2024balancing}. In contrast, underdeveloped governance frameworks lead to accountability gaps and undermine trust in AI-generated outputs, particularly in high-stakes architectural decision making. 

Finally, organizational readiness is key. This encompasses resource availability, technology proficiency, regulatory compatibility (e.g., GDPR compliance), strategic-operational alignment, and the overall coherence between strategic goals and GenAI deployment plans~\cite{alkfairy2025strategic,kaddoumi2022foundational}. Readiness can be categorized into four domains: infrastructure and skilled personnel, data governance maturity, legal and system compatibility, and strategic alignment ~\cite{alkfairy2025strategic}. These collectively contribute to whether GenAI can effectively be integrated into architecture work. In relation to this, research suggests that organizations with greater awareness of agile principles tend to support more adaptive architecture practices, thereby creating more favorable conditions for GenAI integration~\cite{kaddoumi2022foundational}. 

\subsection{Adaptations Required for GenAI Integration}
In response to RQ5, this subsection identifies governance and capability adaptations required to support responsible and effective integration of GenAI in architectural work. 

To support GenAI's potential while mitigating associated risks, organizations should adopt dynamic governance frameworks that evolve with model performance and regulatory change~\cite{Wang2023}. Nakayama et al.'s governance lifecycle proposes an iterative approach that embeds principles such as fairness, transparency, and accountability into GenAI management practices~\cite{nakayama2024balancing}. Architects require well-defined policies for handling sensitive data, interpret model outputs, and apply GenAI tools responsibly. Organizations should also revise data-handling protocols to comply with emerging AI standards and introduce privacy-preserving measures and audit trails to support legal compliance and system transparency~\cite{Alenezi2025, alkfairy2025strategic,Golda2024}. 

Building organizational capabilities remains a priority. Training programs on GenAI risks and best practices can reduce the likelihood of misuse and support cultural acceptance. Employees with strong data-governance awareness are more likely to use GenAI tools in an ethically responsible manner~\cite{amendi2024data}. 

Lastly, trust-building measures, such as open communication and responsible usage guidelines, are key to reducing public skepticism and supporting responsible GenAI adoption within architectural communities~\cite{mahmoud2025public}. 

\section{Discussion}
\label{section:discussion}
This review indicates that GenAI is reshaping roles, workflows, and governance expectations for enterprise architects in agile settings. On the opportunity side, GenAI can support the generation, analysis, and communication of complex information, strengthening the architect’s integrative and decision-making functions within cross-functional teams. On the risk side, limited explainability, contextual misalignment, reliability issues (e.g., hallucinations), and overreliance on automation raise concerns for governance and professional judgement. 

These dynamics suggest a shift from an “architect-as-creator” model toward an “architect-as-curator” role. New competencies, including prompt design, model oversight, and ethical judgement, complement, rather than replace, existing skills. Adoption is not purely technical. Organizations need capability-building, structured evaluation, and support mechanisms, together with governance maturity, ethical frameworks, and cross-team learning. 

A practical implication is tighter integration of GenAI into enterprise architecture platforms to streamline modeling and artifact generation. Theoretically, the findings point to co-adaptive human-AI collaboration. Practically, they emphasize disciplined experimentation, clear data-governance practices, and explicit human-AI task allocation. 

The review remains exploratory. Rapid developments in GenAI limit generalizability. Nonetheless, the review offers guidance for enterprise architects to understand and navigate the implications of GenAI for their evolving roles and practices. 

\section{Conclusion}
\label{section:conclusion}
Based on a structured synthesis of 33 peer-reviewed articles, this SLR concludes that GenAI is influencing the roles, skills, and work practices of architects in agile software development. It augments productivity in ideation, artifact generation (code, diagrams, documentation), and architectural decision support, while introducing risks related to explainability, privacy, and potential erosion of professional judgement through overreliance. Addressing these issues requires targeted education, training, and robust governance. 

The contribution of this review is twofold: it organizes emerging evidence on how architectural roles are evolving with GenAI, and it outlines practical considerations for responsible, context-aware adoption. It also identifies organizational enablers and barriers that inform governance design and strategy for bringing GenAI into agile environments. 

Several limitations apply. Restricting searches to two databases may reduce coverage, although recurring themes across diverse sources suggest stability of the main insights. With many studies being conceptual or exploratory, the review structures early discussions rather than offering definitive causal claims. The term “architects” was used broadly (enterprise, domain, solution, business, IT), which may blur role-specific nuances. Transferability is uncertain and likely depends on enterprise size, maturity, and sector. Finally, the  field of research is nascent and evolving, which constrains the strength of conclusions. 

Future work should pursue empirical studies on the longitudinal effects of GenAI on architectural roles, inter-team collaboration, and enterprise agility. Promising methods include multi-site case studies, ethnographic fieldwork, and controlled experiments. A concrete agenda is to develop principles and patterns that guide architects' effective and responsible use of GenAI. 

\begin{credits}
\subsubsection{\ackname} The authors thank Martin van Capelleveen for the constructive discussions. 

\subsubsection{Disclosure of Interests.}
The authors have no competing interests to declare that are relevant to the content of this article. 
\end{credits}
%
%
%
\bibliographystyle{splncs04}
\bibliography{mybibliography}

\begin{thebibliography}{10}
\providecommand{\url}[1]{\texttt{#1}}
\providecommand{\urlprefix}{URL }
\providecommand{\doi}[1]{https://doi.org/#1}

\bibitem{alkfairy2025strategic}
Al-Kfairy, M.: Strategic integration of generative ai in organizational settings: Applications, challenges and adoption requirements. IEEE Engineering Management Review  (2025). \doi{10.1109/EMR.2025.3534034}

\bibitem{Alenezi2025}
Alenezi, M., Akour, M.: Ai-driven innovations in software engineering: A review of current practices and future directions. Applied Sciences  \textbf{15}(3) (2025). \doi{10.3390/app15031344}

\bibitem{amendi2024data}
Amendi, R., Halim, E., Hartono, H.: {Exploring ethical implications: Unraveling factors influencing data governance awareness behavior in generative AI chatbot}. In: 2024 2nd International Conference on Technology Innovation and Its Applications (ICTIIA). IEEE, Jakarta, Indonesia (2024). \doi{10.1109/ICTIIA61827.2024.10761290}

\bibitem{Waterfall}
Andrei, B.A., Casu-Pop, A.C., Gheorghe, S.C., Boiangiu, C.A.: A study on using waterfall and agile methods in software project management. Journal of Information Systems \& Operations Management pp. 125--135 (2019)

\bibitem{Bahi2024}
Bahi, A., Gharib, J., Youssef, G., Belkadi, L.: {Integrating generative AI for advancing Agile software development and mitigating project management challenges}. International Journal of Advanced Computer Science and Applications  \textbf{15}(3) (2024). \doi{10.14569/IJACSA.2024.0150306}

\bibitem{Bakar2024}
Bakar, N.A.A., Suib, A.H., Othman, A., Amdan, A.A., Hassan, M.A.A., Hussein, S.S.: Artificial intelligence in enterprise architecture: Innovations, integration challenges, and ethics. In: Proceedings of the International Conference on Innovation \& Entrepreneurship in Computing, Engineering \& Science Education (InvENT 2024). pp. 578--588 (2024). \doi{10.2991/978-94-6463-589-8_54}

\bibitem{Banh2023}
Banh, L., Strobel, G.: Generative artificial intelligence. Electronic Markets  \textbf{33}(1), ~63 (2023). \doi{10.1007/s12525-023-00680-1}

\bibitem{beck2001agile}
Beck, K., Beedle, M., van Bennekum, A., Cockburn, A., Cunningham, W., Fowler, M., Grenning, J., Highsmith, J., Hunt, A., Jeffries, R., Kern, J., Marick, B., Martin, R.C., Mellor, S., Schwaber, K., Sutherland, J., Thomas, D.: {Manifesto for Agile software development} (2001), \url{https://agilemanifesto.org/}, accessed: 2025-03-24

\bibitem{Berg2016decision}
van~den Berg, M., van Vliet, H.: The decision-making context influences the role of the enterprise architect. In: Proceedings of the IEEE 20th International Enterprise Distributed Object Computing Workshop (EDOCW). pp.~1--7 (2016). \doi{10.1109/EDOCW.2016.7584389}

\bibitem{vandenberg2016roles}
van~den Berg, M., van Vliet, H.: Enterprise architects should play different roles. In: Proceedings of the IEEE International Conference on Enterprise Distributed Object Computing Workshop (EDOCW). pp.~1--7. IEEE (2016). \doi{10.1109/CBI.2016.56}

\bibitem{NIST2024}
Booth, H., Souppaya, M., Vassilev, A., Ogata, M., Stanley, M., Scarfone, K.: {Secure software development practices for generative AI and dual-use foundation models: An SSDF community profile}. Tech. Rep. NIST Special Publication (SP) 800-218A, National Institute of Standards and Technology (2024). \doi{10.6028/NIST.SP.800-218A}

\bibitem{Breithaupt2021}
Breithaupt, C., Vieracker, J., Chircu, A., Cox, S., Sultanow, E.: {The enterprise architect as a crisis manager: Insights from Lufthansa}. In: INFORMATIK 2020. Lecture Notes in Informatics (LNI), Gesellschaft für Informatik (2021). \doi{10.18420/inf2020_14}

\bibitem{BrightEtAl2024_GenerativeAI_PublicSector}
Bright, J., Enock, F., Esnaashari, S., Francis, J., Hashem, Y., Morgan, D.: {Generative AI is already widespread in the public sector: Evidence from a survey of UK public sector professionals}. Digital Government: Research and Practice  \textbf{6}(1) (2025). \doi{10.1145/3700140}

\bibitem{bughin2024corporate}
Bughin, J.: What drives the corporate payoffs of using generative artificial intelligence? Structural Change and Economic Dynamics  \textbf{71},  658--668 (2024). \doi{10.1016/j.strueco.2024.09.011}

\bibitem{Canat2018}
Canat, M., Català, N.P., Jourkovski, A., Petrov, S., Wellme, M., Lagerström, R.: Enterprise architecture and agile development: Friends or foes? In: 2018 IEEE 22nd International Enterprise Distributed Object Computing Workshop (EDOCW). pp. 176--182. IEEE (2018). \doi{10.1109/EDOCW.2018.00033}

\bibitem{Chahal2019}
Chahal, A., Gulia, P.: Machine learning and deep learning. International Journal of Innovative Technology and Exploring Engineering (IJITEE)  \textbf{8}(12),  4910--4914 (2019). \doi{10.35940/ijitee.L3550.1081219}

\bibitem{McKinsey2023AI}
Chui, M., Yee, L., Hall, B., Singla, A., Sukharevsky, A.: {The state of AI in 2023: Generative AI's breakout year} (2023), \url{https://www.mckinsey.com/capabilities/quantumblack/our-insights/the-state-of-ai-in-2023-generative-ais-breakout-year}, accessed: 2025-03-24

\bibitem{DiRocco2025}
Di~Rocco, J., Di~Ruscio, D., Di~Sipio, C., Nguyen, P.T., Rubei, R.: On the use of large language models in model-driven engineering. Software and Systems Modeling  (2025). \doi{10.1007/s10270-025-01263-8}

\bibitem{gellweiler2019collaboration}
Gellweiler, C.: Collaboration of solution architects and project managers. International Journal of Human Capital and Information Technology Professionals  \textbf{10}(4),  1--15 (2019). \doi{10.4018/IJHCITP.2019100101}

\bibitem{gellweiler2020itarch}
Gellweiler, C.: {Types of IT architects: A content analysis on tasks and skills}. Journal of Theoretical and Applied Electronic Commerce Research  \textbf{15},  15 -- 37 (2020). \doi{10.4067/S0718-18762020000200103}

\bibitem{gellweiler2022alignment}
Gellweiler, C.: {IT architects and IT-business alignment: A theoretical review}. Procedia Computer Science  \textbf{196},  13--20 (2022). \doi{10.1016/j.procs.2021.11.067}

\bibitem{georgiev2024knowledge}
Georgiev, D., Antonova, A.: {Enhancing knowledge sharing processes via automated software documentation management systems using Gen AI software tools}. In: Proceedings of the IEEE Conference. IEEE (2024). \doi{10.1109/MMA62616.2024.10817681}

\bibitem{Golda2024}
Golda, A., Mekonen, K., Pandey, A., Singh, A., Hassija, V., Chamola, V., Sikdar, B.: Privacy and security concerns in generative ai: A comprehensive survey. IEEE Access  \textbf{12},  48126--48149 (2024). \doi{10.1109/ACCESS.2024.3381611}

\bibitem{TOGAF9_2025}
Group, T.O.: Togaf 9.2 documentation (2025), \url{https://pubs.opengroup.org/architecture/togaf9-doc/arch/index.html}

\bibitem{Guo2021}
Guo, H., Smite, D., Li, J., Gao, S.: Enterprise architecture and agility: A systematic mapping study. In: Business Modeling and Software Design (BMSD 2021), Lecture Notes in Business Information Processing. vol.~422, pp. 296--305. Springer (2021). \doi{10.1007/978-3-030-79976-2_18}

\bibitem{Hagos2024}
Hagos, D.H., Battle, R., Rawat, D.B.: Recent advances in generative ai and large language models: Current status, challenges, and perspectives. IEEE Transactions on Artificial Intelligence  \textbf{5}(1) (2024). \doi{10.1109/TAI.2024.3444742}

\bibitem{Hamza2024}
Hamza, M., Siemon, D., Akbar, M.A., Rahman, T.: Human-ai collaboration in software engineering: Lessons learned from a hands-on workshop. In: ACM/IEEE International Workshop on Software-intensive Business (2024). \doi{10.1145/3643690.3648236}

\bibitem{Handler2024}
Handler, A., Larsen, K.R., Hackathorn, R.: Large language models present new questions for decision support. International Journal of Information Management  \textbf{79} (2024). \doi{10.1016/j.ijinfomgt.2024.102811}

\bibitem{kaddoumi2022foundational}
Kaddoumi, T., Watfa, M.: A foundational framework for agile enterprise architecture. International Journal of Lean Six Sigma  \textbf{13}(1),  136--155 (2022). \doi{10.1108/IJLSS-03-2021-0057}

\bibitem{kalenda2018scaling}
Kalenda, M., Hyna, P., Rossi, B.: Scaling agile in large organizations: Practices, challenges, and success factors. Journal of Software: Evolution and Process  \textbf{30}(10),  e1954 (2018). \doi{10.1002/smr.1954}

\bibitem{kempegowda2020essential}
Kempegowda, S.M., Chaczko, Z.: Essential skill of enterprise architect practitioners for digital era. In: Proceedings of the International Conference on Electrical and Data Engineering. IEEE (2020). \doi{10.1109/ICSENG.2018.8638210}

\bibitem{kitchenham2018guidelines}
Kitchenham, B., Charters, S.: Guidelines for performing systematic literature reviews in software engineering  (2007)

\bibitem{krishnamurthy2017solution}
Krishnamurthy, R.: Breezing my way as a solution architect: A retrospective on skill development and use. IEEE Software  \textbf{34}(3),  9--13 (2017), \url{https://doi.org/10.1109/MS.2017.83}

\bibitem{kurnia2021enterprise}
Kurnia, S., Kotusev, S., Shanks, G., Dilnutt, R., Taylor, P., Milton, S.K.: Enterprise architecture practice under a magnifying glass: Linking artifacts, activities, benefits, and blockers. Communications of the Association for Information Systems  \textbf{49},  668--698 (2021). \doi{10.17705/1CAIS.04936}

\bibitem{Lankhorst2017}
Lankhorst, M.: Enterprise architecture at work: Modelling, communication and analysis. The Enterprise Engineering Series, Springer (2017). \doi{10.1007/978-3-662-53933-0}

\bibitem{mahmoud2025public}
Mahmoud, A.B., Kumar, V., Spyropoulou, S.: Identifying the public's beliefs about generative artificial intelligence: A big data approach. IEEE Transactions on Engineering Management  (2025). \doi{10.1109/TEM.2025.3534088}

\bibitem{mulder2023modern}
Mulder, J.: Modern enterprise architecture: Using DevSecOps and cloud-native in large enterprises. Apress (2023). \doi{10.1007/978-1-4842-9066-8}

\bibitem{nafei2016organizational}
Nafei, W.: Organizational agility: The key to improve organizational performance. International Business Research  \textbf{9}(3),  97--104 (2016). \doi{10.5539/ibr.v9n3p97}

\bibitem{Nah2023}
Nah, F.F.H., Zheng, R., Cai, J., Siau, K., Chen, L.: {Generative AI and ChatGPT: Applications, challenges, and AI-human collaboration}. Journal of Information Technology Case and Application Research  \textbf{25}(3),  277--304 (2023). \doi{10.1080/15228053.2023.2233814}

\bibitem{nakayama2024balancing}
Nakayama, M., Wan, Y., Alvino, F.: {Balancing potential and prudence: An on-going study on organizational policymaking for generative AI adoption}. In: Proceedings of the 2024 Artificial Intelligence for Business (AIxB). IEEE (2024). \doi{10.1109/AIxB62249.2024.00014}

\bibitem{nidhisree2024generative}
Nidhisree, C., Paul, A., Venunadh, A., Bhowmick, R.S.: {Generative AI under scrutiny: Assessing the risks and challenges in diverse domains}. In: 2024 IEEE 6th International Conference on Cybernetics, Cognition and Machine Learning Applications (ICCCMLA). pp. 243--246. IEEE (2024). \doi{10.1109/ICCCMLA63077.2024.10871350}

\bibitem{Norheim2024}
Norheim, J.J., Rebentisch, E., Xiao, D., Draeger, L., Kerbrat, A., de~Weck, O.L.: Challenges in applying large language models to requirements engineering tasks. Design Science Journal  \textbf{10} (2024). \doi{10.1017/dsj.2024.8}

\bibitem{PRISMA2021}
Page, M.J., McKenzie, J.E., Bossuyt, P.M., Boutron, I., Hoffmann, T.C., Mulrow, C.D., Shamseer, L., Tetzlaff, J.M., Akl, E.A., Brennan, S.E., Chou, R., Glanville, J., Grimshaw, J.M., Hróbjartsson, A., Lalu, M.M., Li, T., Loder, E.W., Meerpohl, J.J., Munn, S., McGuinness, L.A., Stewart, E.S., Thomas, J., Tricco, A.C., Barbour, V., Garner, P.C., Ioannidis, J.P., Moher, D.: {The PRISMA 2020 statement: An updated guideline for reporting systematic reviews}. BMJ  \textbf{372}, ~n71 (2021). \doi{10.1136/bmj.n71}

\bibitem{Paliwal2024}
Paliwal, G., Donvir, A., Gujar, P., Panyam, S.: {Accelerating time-to-market: The role of generative AI in product development} (2024). \doi{10.1109/COLCOM62950.2024.10720255}, white paper or preprint

\bibitem{OSF}
Piest, J.P.S., Kooy, S.J., Bemthuis, R.: {Impact and implications of GenAI on enterprise architects} (2025). \doi{10.17605/OSF.IO/DH2AS}

\bibitem{putta2018adopting}
Putta, A., Paasivaara, M., Lassenius, C.: {Adopting Scaled Agile Framework (SAFe): A multivocal literature review}. In: XP '18 Companion. ACM (2018). \doi{10.1145/3234152.3234164}

\bibitem{rashid2024ai}
Rashid, A.B., Kausik, A.K.: {AI revolutionizing industries worldwide: A comprehensive overview of its diverse applications}. Hybrid Advances  \textbf{7}(100277),  100277--100277 (2024). \doi{10.1016/j.hybadv.2024.100277}

\bibitem{razavian2023odaks}
Razavian, M., Paech, B., Tang, A.: The vision of on-demand architectural knowledge systems as a decision-making companion. Journal of Systems and Software  \textbf{198} (2023). \doi{10.1016/j.jss.2022.111560}

\bibitem{Rouhani2015}
Rouhani, B.D., Mahrin, M.N.M., Nikpay, F., Ahmad, R., Nikfard, P.: A systematic literature review on enterprise architecture implementation methodologies. Information and Software Technology  \textbf{62},  1--20 (2015). \doi{10.1016/j.infsof.2015.01.012}

\bibitem{sassa2023scrum}
Sassa, A.C., de~Almeida, I.A., Fernandes~Pereira, T.N., de~Oliveira, M.S.: Scrum: A systematic literature review. International Journal of Advanced Computer Science and Applications  \textbf{14}(4),  173--184 (2023). \doi{10.14569/IJACSA.2023.0140420}

\bibitem{scaledagile_enterprise_architect}
{Scaled Agile, Inc.}: Enterprise architect (2023), \url{https://scaledagileframework.com/enterprise-architect/}, accessed: 2025-03-24

\bibitem{scaledagile_solution_architect}
{Scaled Agile, Inc.}: Solution architect (2023), \url{https://scaledagileframework.com/solution-architect/}, accessed: 2025-03-24

\bibitem{scaledagile_system_architect}
{Scaled Agile, Inc.}: System architect/engineer (2023), \url{https://scaledagileframework.com/system-architect-engineer/}, accessed: 2025-03-24

\bibitem{strano2007enterprise}
Strano, C., Rehmani, Q.: The role of the enterprise architect. Information Systems and e-Business Management  \textbf{5}(4),  379--396 (2007). \doi{10.1007/s10257-007-0053-1}

\bibitem{ullrich2021vuca}
Ullrich, A., Bertheau, C., Wiedmann, M., Sultanow, E., Körppen, T., Bente, S.: Roles, tasks and skills of the enterprise architect in the vuca world. In: 2021 IEEE 25th International Enterprise Distributed Object Computing Workshop (EDOCW). pp. 261--270 (2021). \doi{10.1109/EDOCW52865.2021.00057}

\bibitem{uludag2019expectations}
Ömer Uludağ, Kleehaus, M., Matthes, F., Schneider, J.: {What to expect from enterprise architects in large-scale Agile development: A multiple-case study}. In: Proceedings of the Americas Conference on Information Systems (AMCIS) (2019)

\bibitem{uludag2020supporting}
Ömer Uludağ, Matthes, F.: Investigating the role of enterprise architects in supporting large-scale agile transformations: A multiple-case study. In: Proceedings of the Americas Conference on Information Systems (AMCIS). Salt Lake City, USA (2020), \url{https://aisel.aisnet.org/amcis2020/digital_agility/digital_agility/2/}

\bibitem{uludag2020patterns}
Ömer Uludağ, Matthes, F.: Large-scale agile development patterns for enterprise and solution architects. In: Proceedings of the European Conference on Pattern Languages of Programs (EuroPLoP). ACM (2020). \doi{10.1145/3424771.3424895}

\bibitem{Wagner2016}
Wagner, H.T., Moshtaf, J.: Individual it roles in business–it alignment and it governance. In: Proceedings of the 2016 European Conference on Information Systems (ECIS) (2016). \doi{10.1109/HICSS.2016.610}

\bibitem{Wang2023}
Wang, Y., Pan, Y., Yan, M., Su, Z., Luan, T.H.: {A survey on ChatGPT: AI–generated contents, challenges, and solutions}. IEEE Open Journal of the Computer Society  (2023). \doi{10.1109/OJCS.2023.3300321}

\bibitem{waterman2015agile}
Waterman, M., Noble, J., Allan, G.: {How much up-front? A grounded theory of Agile architecture}. In: IEEE Xplore (2015). \doi{10.1109/ICSE.2015.54}

\bibitem{williams1983aiintroduction}
Williams, C.: A brief introduction to artificial intelligence. In: Proceedings OCEANS '83. pp. 94--99 (1983). \doi{10.1109/OCEANS.1983.1152096}

\bibitem{Zachman1987}
Zachman, J.A.: A framework for information systems architecture. IBM Systems Journal  \textbf{26}(3),  276--292 (1987). \doi{10.1147/sj.263.0276}

\bibitem{Zachman2000}
Zachman, J.A.: The framework for enterprise architecture: Background, description and utility (2000), \url{https://zachman-feac.com/the-framework-for-enterprise-architecture-background-description-and-utility}

\end{thebibliography}

\end{document}